\documentclass{emulateapj}

\shorttitle{Overdensities at $z \sim 3.7$ in the $Chandra$ deep field-south}
\shortauthors{Kang, Im}

\begin{document}
\title{Overdensities of galaxies at $z \sim 3.7$ in $Chandra$ Deep Field-South}
\author{Eugene Kang\altaffilmark{1}, Myungshin Im\altaffilmark{1}} 
\altaffiltext{1}{CEOU, Astronomy Program, Department of Physics and Astronomy, 
Seoul National University, 56-1 Shillim 9-dong, Kwanak-gu, Seoul 151-742, Korea}

\begin{abstract}
  We report the discovery of possible overdensities of galaxies at 
 $z \sim 3.7$ in $Chandra$ Deep Field South (CDF-S). 
  These overdensities are identified from a photometric redshift-selected
 sample, and the $BVz$-selected sample.
  One overdensity is identified in the proximity of two active galactic nuclei
 and Lyman break galaxies
 at $z=3.66$ and $z=3.70$ at 7$\sigma$ significance level.
  The other overdensity is less significant. It is identified around
 six $z_{spec} \simeq 3.6$ galaxies at 3$\sigma$ significance level.
  The line-of-sight velocity dispersions of these overdensities are found to be
 $\sigma_{v} \simeq 500 - 800$ km s$^{-1}$, comparable to the velocity dispersions
 of clusters of galaxies today.
  Through spectral energy distribution fitting,
 we find $\sim$ 15 massive galaxies with $M \gtrsim 10^{11}\,M_{\odot}$
 around the $z \simeq 3.7$ overdensity.
  The mass of the $z \simeq 3.7$ overdensity is found to be 
 a few $\times 10^{14}\,M_{\odot}$.
  Our result suggests that high-redshift overdense regions
 can be found in a supposedly blank field,
 and that the emergence of massive structures can be traced back to redshifts
 as high as $z \sim 3.7$.  
\end{abstract}


\section{Introduction}
  In hierarchical galaxy formation, galaxy clusters
 grow through gravitational attraction
 of matter around high $\sigma$ peaks, and they are predicted to be  
 rare at high redshift. Therefore, constraining
 the number density of high-redshift clusters is an important method of testing 
 hierarchical models (Evrard et al. 2002; Mantz et al. 2008).

  Many previous studies have been carried out to search for 
 evolved structures at high redshift.
  Radio galaxies and active galactic nuclei (AGNs) are considered to be signposts 
 for high-redshift proto-clusters, and several studies have found
 proto-clusters at $2 \lesssim z \lesssim 4$
 \citep{lef96,kur00,pen00,wol03,kaj06,kod07,ven07}. 
 Overdense regions are also identified out to $z \sim 6$ 
 using galaxies selected with the Lyman break 
 technique or narrowband imaging (e.g, Ouchi et al. 2005).

  When identifying overdense regions, it is necessary to compare the
 overdensity with a control field which lacks notable high-redshift
 overdense regions. For such a purpose, many studies have used 
 the $Chandra$ Deep Field South (CDF-S) due to the wealth
 of the deep multiwavelength data over a moderately
 large field of view \citep{kaj06,kod07}.
  However, a matter of concern is whether the CDF-S can really be considered
 as a field devoid of high-redshift overdense regions.
  There exist a fair number of high-redshift X-ray-detected 
 AGNs in the CDF-S, which may harbor high-redshift clusters.

  Motivated by this, we searched for signs of overdensity around
 two AGNs at $z=3.66$ and $3.70$ separated by 
 $2\arcmin.7$ in the CDF-S, and we report the discovery of
 possible overdense regions at $z \sim$ $3.7$ in the CDF-S.  
   
   Throughout this Letter,
 we assume a cosmology with nonzero cosmological constant
 (Im et al. 1997), $\Omega_m = 0.3$, $\Omega_\Lambda = 0.7$ and $h = 0.7$,
 consistent with the $WMAP$ cosmology \citep{spe07}. All magnitudes are given
 in the AB system.
 
\section{Sample selection of high redshift galaxy}
  We used two galaxy samples to identify overdense areas
 at $z=$3.6-3.7; a $BVz$-selected sample, and galaxies selected with
 photometric redshifts.

\subsection{$BVz$ Selection}

  The Lyman break technique has proved to be 
 an efficient method to select galaxies at high redshift
 from multicolor optical data (Steidel et al. 1996).
 For example, Giavalisco et al. (2004b) selected B-band dropouts
 at  $z \sim 4$ from GOODS data.
 However, since the $B_{435}$-dropout selection contains galaxies
 over a broad redshift range, we defined a color space optimized
 to select galaxies at $z \simeq$3.6-3.7 using 
 available spectroscopic redshifts including 17 at 
 $z \simeq$3.6-3.7 (Vanzella et al. 2006, 2008; Popesso et al. 2008).
 Figure 1 shows our selection box, overlaid on galaxies in the
 CDF-S field. The figure demonstrates that our selection criteria
 effectively filter out objects at $z < 3.55$, and at $z > 3.75$.
  For example, there are 18 galaxies at $z_{spec} \sim 3.47$, but only
 one of them makes it into the $BVz$-selected sample.
  For the photometric data, we used the version r1.1 
 Advanced Camera for Surveys (ACS) multiband catalog
 of the GOODS team (Giavalisco et al. 2004a), and 
 the magnitude cut of $z_{850} \leq 26.5$ mag was chosen. 

  With this method, we selected 245 objects over the CDF-S area
 of $\sim$160 arcmin$^2$ corresponding to a 
 surface density of $\Sigma = 1.53 \pm 0.06$ arcmin$^{-2}$ 
 (Poisson noise). 

\subsection{Photometric redshifts}
 
  The $BVz$ selection is biased for UV-bright galaxies. As a complementary
 sample, we also selected $K$-band-limited (${K_s} \leq 23.8$ mag),
 $z \sim$3.6-3.7 galaxy candidates using photometric redshift.

  We estimated the photometric redshifts of the 
 $K_s$-band-limited objects using the Bayesian photometric redshift estimation
 (BPZ; Ben\'{i}tez 2000).
  The photometric data include \\$UU^\prime$$B_{435}V_{606}i_{775}z_{850}JHK$ bands,
 and $Spitzer$ IRAC photometry (Giavalisco et al. 2004a).
  The Near-infrared (NIR) photometry was performed using the version 2.0 released images
 in the $J$-, $H$-, and $K_{s}$-bands of the ESO-GOODS team
 (J. Retzlaff et al. 2008, in preparation).
 The depth of the VLT/ISAAC data reaches $K=24.7$ at the 5$\sigma$
 AB limits, and the simultaneous coverage in $JHK_s$ bands
 is 160 arcmin$^2$ after trimming the edges.
 To create a catalog, we ran SExtractor (Bertin $\&$ Arnouts 1996)
 in double-image mode, using $K_{s}$-band detections as a reference,
 and performing photometry in the $J$- and $H$-bands at
 the $K_{s}$ positions.
  Auto-magnitudes were taken for the $J$,$H$, and $K_{s}$ photometry.
  The IRAC photometry catalog was kindly provided by R. Chary
  (2008, private communication). In most cases,
   we found IRAC detections for the $K$-band selected objects.

  To derive photometric redshifts, we used 742 spectral energy distribution (SED) templates with 
 a single burst, or a constant star-formation rate (SFR), or an
 exponentially declining SFR in the form of SFR $\propto$ exp(t/$\tau$) for 
 $\tau$ =  1, 2, 3, 5, 15, or 30Gyr. 
  These template SEDs were generated using a stellar population synthesis model
 of Bruzual \& Charlot (2003). 
  The metallicity was varied to be $0.4 Z_{\odot}$, $1 Z_{\odot}$, or $2 Z_{\odot}$,
 and we adopted the Salpeter initial mass function extending from 0.1 to
 100 $M_{\odot}$.  The SED ages range from 100 Myr to 10 Gyr or 
 to the age of the universe at the corresponding redshift.
 The reddening parameter $A_{V}$ was allowed to change from 0 to 1.8
 according to the simple two-component model of Charlot \& Fall (2000). 
\\ From the comparison of $z_{spec}$ versus $z_{phot}$, 
 we find that contamination from low-redshift interlopers is 
 minimal when selecting $z \sim$3.6-3.7 objects, 
 at the redshift interval of $3.45 < z_{phot} < 3.85$.
 When we restrict the redshift interval as above, the scatter in
 the $z_{phot}$ versus $z_{spec}$ relation is about 0.1 after excluding
 several clear outliers.
  Therefore, we choose $3.45 < z_{phot} < 3.85$ as the interval for
 the photo-$z$ sample selection.
  This gives 60 objects in total, including 17 spectroscopically confirmed
 galaxies  with good $z_{spec}$ quality flags (B or better) at $z \simeq$3.6-3.7.
  Note that we use $z_{spec}$ instead of $z_{phot}$ when available, but 
 restrict the spectroscopic redshift range to be $3.585 < z < 3.706$.
  With the photo-$z$ selection criteria, 73\% of $z_{spec} =$3.6-3.7
 galaxies are detected as $z=$3.6-3.7 galaxy candidates. 
  The contamination rate is 26\%, but the contamination due to low redshift objects
 ($z_{spec} < 3.3$) is zero. Most of the contamination comes from
 $z_{spec} \sim 3.5$ objects.

\begin{figure}[bht]
\centering
\resizebox*{3in}{3in}{\includegraphics{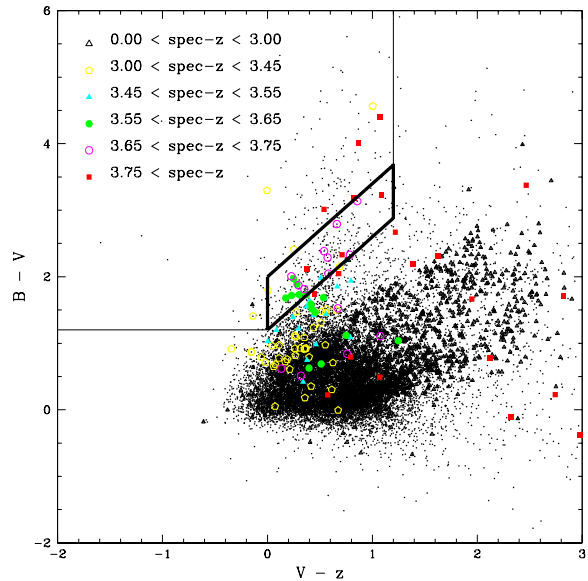}}
\caption{$V-z$ vs. $B-V$ 
 colors of objects in the CDF-S with $z < 26.5$ mag. The parallelogram in the middle of the figure 
 represents the selection criteria of our $BVz$-selected sample. This is similar to the $B$-band
 dropout selection box (thin line; Giavalisco et al. 2004b).
 The spectroscopic data are overplotted.}%}}
\end{figure}

\begin{figure*}[bht]
\centering
       \hspace{-2.05in}
\resizebox*{2.3in}{2.3in}{\includegraphics{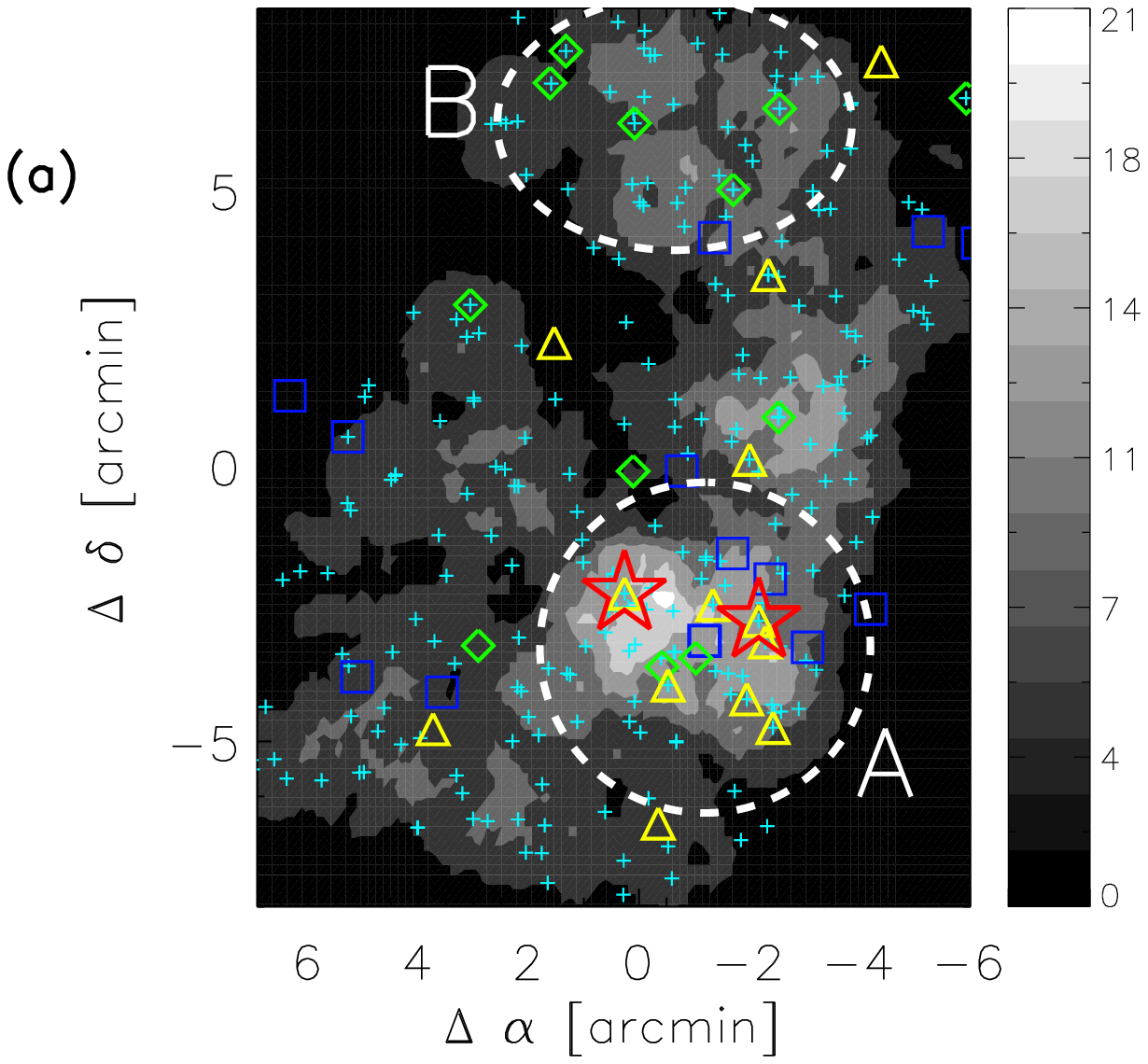}}
       \hspace{-0.1in}
\resizebox*{2.0in}{2.3in}{\includegraphics{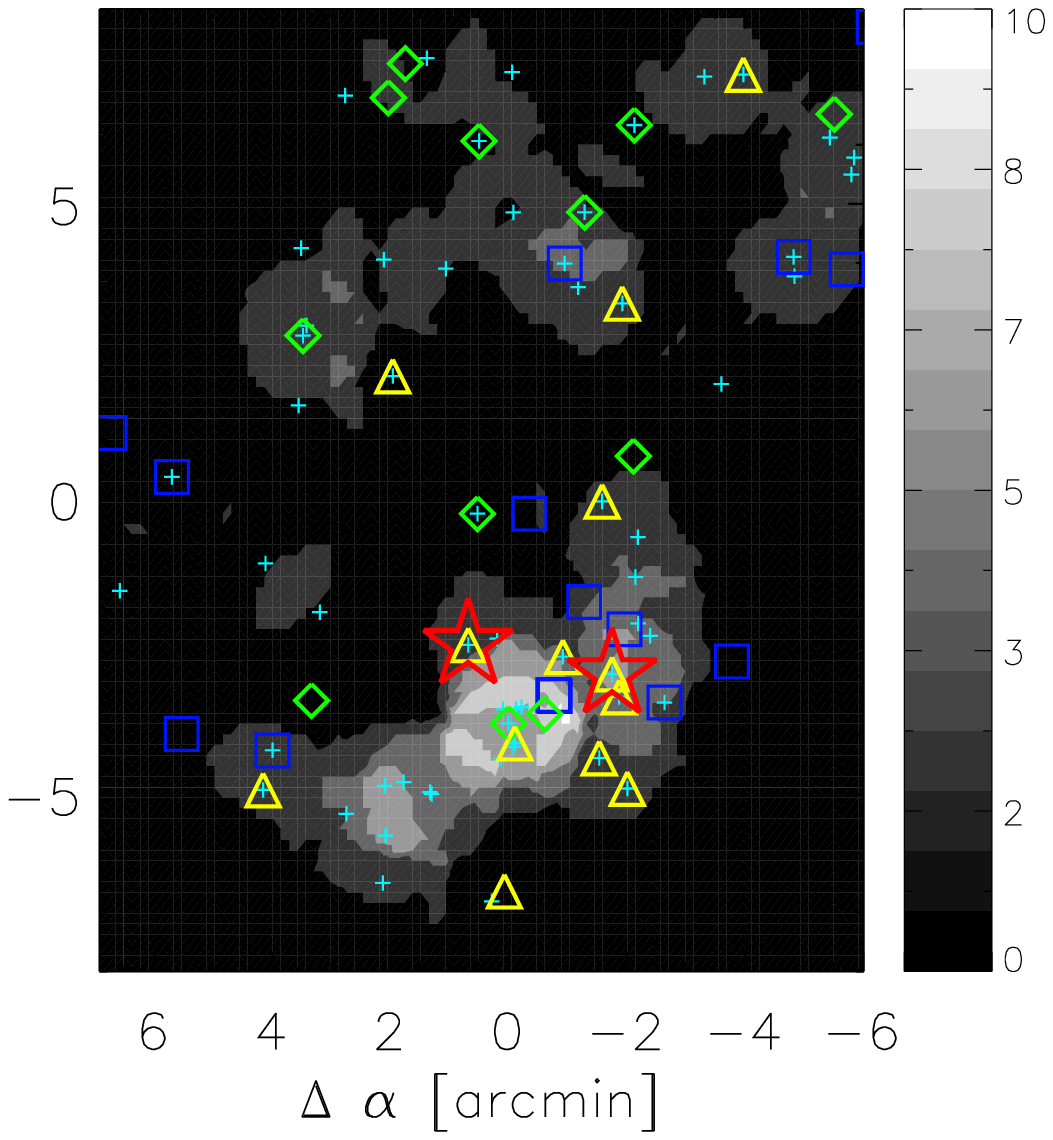}}
        \hspace{0.1in}
\resizebox*{2.5in}{2.0in}{\includegraphics{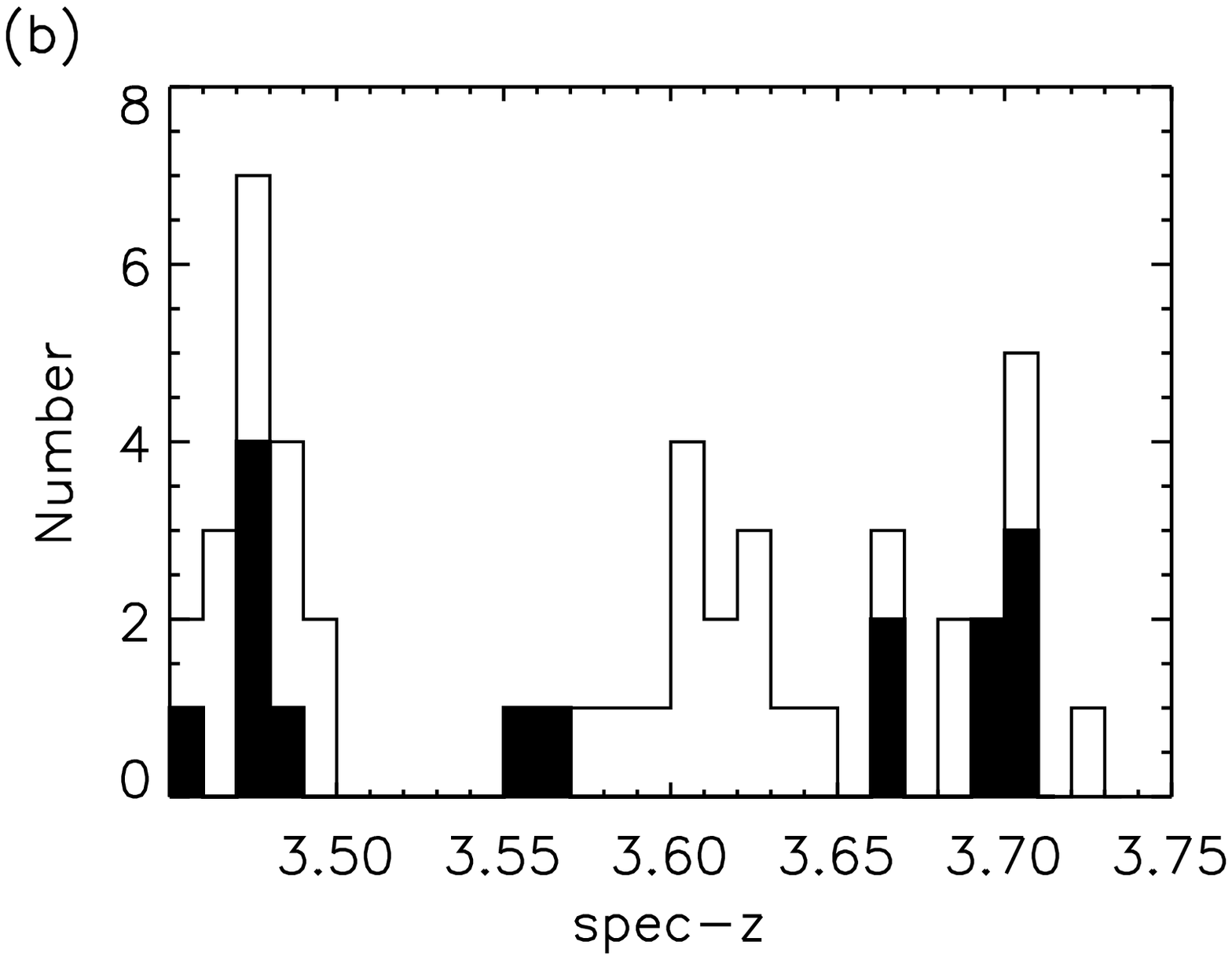}}
        \hspace{-1.9in}
%\caption{\fontfamily{ptm}\selectfont{\normalsize{(a, $Left$):
\caption {(a, Left): Spatial distribution of the $BVz$-selected sample (far-left)
 and photo-$z$ selected objects, both indicated by crosses.
 Squares: $3.45 \leq z_{spec} < 3.55$; diamonds: $3.55 \leq z_{spec} < 3.65$;
 triangles: $3.65 \leq z_{spec} < 3.75$; stars: AGNs at $z_{spec}$= 3.66
and 3.70. The possible overdensities are identified with dashed circles, and
their peak locations are at ${03^h}{32^m}{26^s},-27\degr{51^m}{06^s}$ (J2000)
for the $BVz$-selected sample, and ${03^h}{32^m}{24^s},-27\degr{50^m}{50^s}$,
for the photo-$z$ selected objects.
(b, Right): Redshift distribution for the spectroscopic sample at $3.45 < z < 3.75$.
  The shaded region shows the objects located within 1.0 Mpc radius
 of the southern overdensity peak, A.}
\end{figure*}

\begin{figure*}[bht]
      \hspace{0.7in}
\resizebox*{2.2in}{2.5in}{\includegraphics{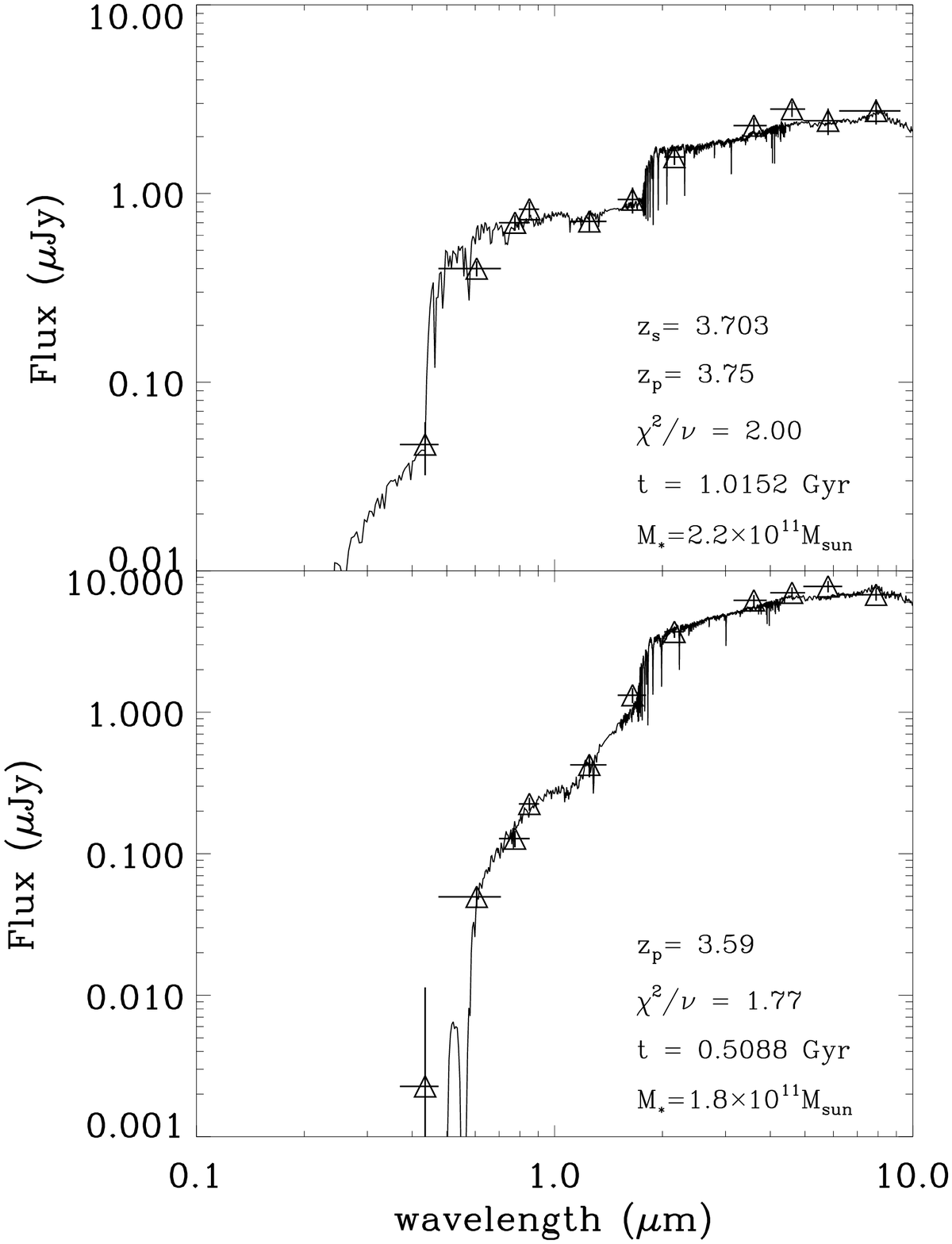}}
      \hspace{0.7in}
\resizebox*{2.4in}{2.4in}{\includegraphics{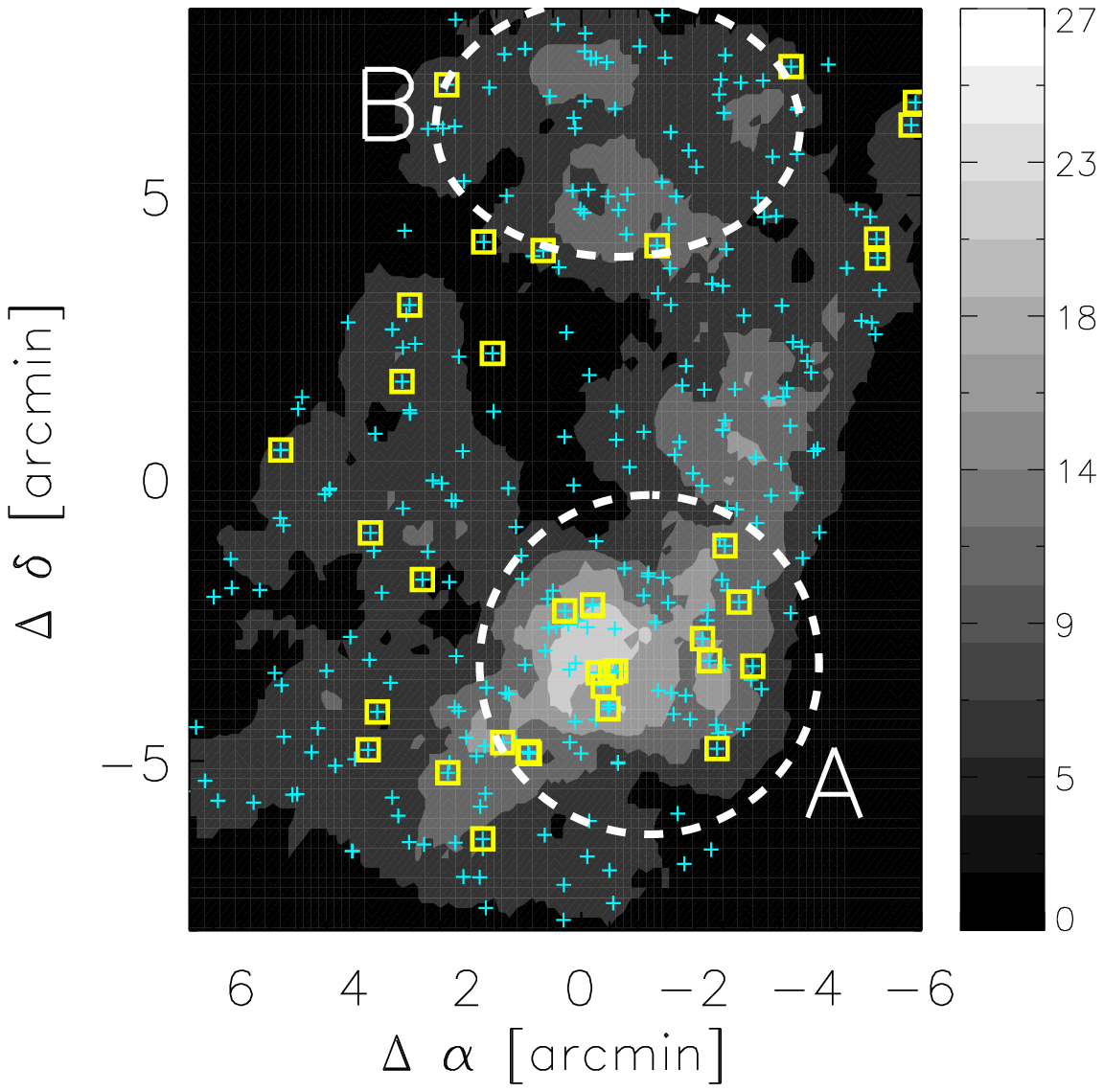}}
%\caption{\fontfamily{ptm}\selectfont{\normalsize
\caption{Massive galaxies ($M_{star} >  10^{11}\,M_{\odot}$)
in the CDF-S. Left: The optical to 8$\micron$
SEDs for two representative massive galaxies
 associated with the overdense region A. The best-fit Bruzual \& Charlot (2003) models
to the template SEDs (solid line) are shown.
Right: the spatial distribution of massive galaxies (square)
 on the sum of two samples (cross) peaked at ${03^h}{32^m}{23^s},-27\degr{52^m}{19^s}$.}%}}}

\end{figure*}

\section{Results}

 We searched for the overdensities by running circular top-hat 
 filters with various diameters ($0\arcmin.5$ to $4\arcmin.8$, corresponding
 to 0.2-2 Mpc at $z=3.7$) on the two-dimensional spatial 
 distribution of galaxies in the above samples. The number of
 objects was counted around a given point within a circle
 with a particular diameter, producing a density contour map. 
   Figure 2(a) shows the surface number density contour map
 and the location of $BVz$-selected sample and
 photo-$z$ selected objects, respectively.
  Here, the contours are created using $2\arcmin.4$ diameter filters,
 which corresponds to $\sim$1.0 Mpc in physical size at $z=3.7$.
  The 1 Mpc size is found to be the optimal size to identify 
 the $z=3.7$ overdensity (see below).
  The spectroscopically confirmed galaxies are plotted with squares
 ($3.45 \leq z_{spec} < 3.55$), diamonds ($3.55 \leq z_{spec} <3.65$),
 and triangles ($3.65 \leq z_{spec} < 3.75$), and the open stars 
 represent AGNs at $z_{spec} =$3.66 and 3.70 from Szokoly et al. (2004).
  One density peak is apparent, located in the southern part of 
 the CDF-S (A in Figure 2(a)), around $z=3.66$, $z=3.70$ AGNs,
 and $\langle z_{spec}\rangle$=3.70 Lyman break galaxies (LBGs, diamonds).
  Another possible peak exists in the north around $\langle z_{spec}\rangle=$
 3.60 LBGs (B), but its significance is much weaker.

\begin{deluxetable}{ccccc}
\tabletypesize{\footnotesize}
\tablewidth{0pc}
\tablenum{1}
\tablecaption{Significance of overdensity}
\tablehead{ \colhead{sample} & \colhead{mean density} & \colhead{1-$\sigma$} &\colhead{A (South)} & \colhead{B (North)}
\nl \colhead{} & \colhead{arcmin$^{-2}$} & \colhead{} & \colhead{deviation $\sigma$} & \colhead{deviation $\sigma$}}
\startdata
\nl $BVz$-selected sample & 1.48 & 0.64 & 4.94 & 2.52 \\
    $3.45 \le z_{phot} \le 3.85$ & 0.37 & 0.27 & 6.86 &1.96 \\
    Sum of two samples& 1.68 & 0.69 & 6.18 & 2.37 \\
\enddata
\end{deluxetable}

  To quantify the statistical significance of these peaks,
 we measured the overdensity factor as a number of $\sigma$ from 
 the mean surface number density by performing a Gaussian fit 
 over the surface number density histogram.
  The significances of the overdensities are summarized in Table 1.  
  The significance of the overdensity is $5-7\sigma$ for the A overdensity,
 demonstrating that its significance is very strong.
  The significance of the overdensity does not change for  
 the top-hat filter size between 0.75 and 1.3 Mpc. Beyond these sizes,
 the significance decreases.
 On the other hand, the B overdensity has a significance of 
 only $2-3\sigma$.

  With the limited number of spectroscopic redshifts,
 assigning redshifts of the overdense regions is not an easy task.
  Nevertheless, we find several redshifts to be plausible.
  The redshift distribution of CDF-S galaxies with $z_{spec}$ shows
 three distinct peaks at $z_{spec} \sim 3.47$, $z_{spec} \sim
 3.60$, and $z_{spec} \sim 3.70$ (Figure 2(b)).
  The majority of $z_{spec} \sim 3.70$ objects are located around
 the overdense area A (filled histogram),
 as well as two AGNs at $z_{spec} = 3.70$ and $3.66$,
 suggesting strongly that this overdensity is at $z \simeq 3.7$.
  There are also several
 $z_{spec} \sim 3.47$ galaxies near the A overdensity. However,
 the overdensity is more likely to be at $z \sim 3.7$,
 since the overdense area is clearly identified in the contour map of
 the $BVz$ sample which effectively filters out $z < 3.55$ galaxies.

  Around the weak, overdense region B, we find six LBGs
 with $\langle z \rangle =3.60$, and find a one-dimensional velocity dispersion
 of $\sigma_{v} \simeq 500$ km s$^{-1}$ for these objects.
  The B overdense region may be a loosely bound overdensity at $z = 3.6$.

\section{Discussion}

\subsection{Cosmic Variance}

  We examined whether the number of high-redshift candidates is
 abnormally high in this field compared to other fields.
 For this, we examined only the $BVz$-selected sample,
 since the photo-$z$ sample does not
 have comparative samples readily available.

  We studied the number density of $BVz$-selected objects
 in the GOODS-North field to compare with the CDF-S.
 Using the version r1.1 ACS multiband source catalog,
 we find 220 $BVz$-selected objects in the GOODS-North field
 (see Giavalisco et al. 2004b).
 This gives the surface density $\Sigma$ = 1.38 $\pm$ 0.07 arcmin$^{-2}$,
 comparable to the number in CDF-S ($\Sigma = 1.53 \pm 0.06$ arcmin$^{-2}$).

  We also looked for a sign of overdensity at $z \sim 3.7$
 using the $BVz$-selected sample in the GOODS-North field. We find the strongest
 overdensity has the overdensity significance of $3.7\sigma$.
  This suggests that the GOODS-North may contain a weak overdense area at $z \sim 3.7$, but
 its significance is much lower than the overdense region
 we identified in the CDF-S.

\subsection{Stellar Population}

  In order to find massive galaxies associated with the overdense regions,
  we performed SED fittings to 60 galaxies in the photo-$z$ sample following
  the procedure of Shim et al. (2007), 
  with the SED templates used for the photometric redshift 
  determination.
   Note that we have limited the SED fitting to objects with IRAC data
 in order to reduce uncertainties and degeneracies arising from
 using only the rest-frame optical data.

  In Figure 3, we show the best-fit SEDs of two representative
 massive galaxies at the overdense region A in the CDF-S.
  We find 35 galaxies with $M_{star} \gtrsim  10^{11}\,M_{\odot}$
 over the whole CDF-S area.
 More than 43\% (15) of these massive objects
 are located within $\sim$ 1 Mpc radius of 
 the overdensity peak of the A region.
 This offers a further support for the existence of $z \sim 3.7$ overdensity.

  We also checked the usefulness of the $H-K$ color selection of 
 $z \sim 3.7$ as advocated by Brammer \& van Dokkum (2007). They suggest
 that objects with $H-K > 0.9$ mag are likely to be at 
 $\langle z \rangle = 3.7$ based on their photometric redshift analysis.
  However, our analysis shows that only about 30\% of $H-K > 0.9$ objects 
 are at $z > 3$, for both the $z_{spec}$ and the $z_{phot}$ samples.
  Therefore, we conclude that the usefulness of the $H-K > 0.9$ selection
  method is limited.

\subsection{Overdensity Mass}

  We estimate the overdensity mass in several different ways for the A area. 

  First, we assume that it is a virialized
 structure, following Biviano et al. (2006).
  This assumption is not plausible, considering that 
 the age of the universe is only 1.6 Gyr at $z=3.7$, while it takes nearly
 several Gyr for galaxies to cross the 1 Mpc structure at the given 
 velocity dispersion. Nevertheless, the method can place a useful upper
 limit on the overdensity mass.  
  For the cluster size, we use the harmonic mean radius of the photo-$z$ selected
 objects within a 1 Mpc radius (twice the filter size that gives
 the maximal overdensity signal) from the overdensity center. This gives a 
 size of 0.55 Mpc. 
  Assuming that the one-dimensional velocity dispersion, $\sigma_v$, 
 of several possible proto-cluster members with $z_{spec}$ is a representative
 value,  we estimate the cluster virial mass using the gapper estimator
 (Beers et al. 1990).
 We find that $\sigma_v =$ 800 km s$^{-1}$ by excluding a small redshift peak
 at $z_{spec}=3.66$.
 The derived cluster virial mass is,
 then $M_{vir} \simeq 5 \times 10^{14}\,M_{\odot}$.

  Next, we apply a 
 procedure similar to that of Venemans et al. (2005) on
 the $z_{phot}$-selected sample.
  Suppose that the cluster is roughly spherical, and has a 0.55 Mpc radius 
 equivalent to the harmonic mean radius as derived above.
  Using the range $3.45 < z < 3.85$
 as the selection window, we find 
 an overdensity factor of $\delta_{gal} \sim 700$.
 Adopting the bias parameter of $b=4$ of B-dropouts in the CDF-S (Lee et al.
 2006),
 we obtain a cluster mass of $4 \times 10^{14}$  $M_{\odot}$.

  Finally, we add up the stellar mass
 of the member candidates within the 0.55 Mpc radius,
 and convert it to the total mass assuming a 
 $M_{star}$-to-$M_{halo}$ ratio of 0.026 - 0.056 of early-type
 galaxies (Jiang \& Kochanek 2007). 
  Here, the overdensity mass can be expressed as
 $M=(N_{g} \times f \times \langle M_{star}\rangle)/(0.026 - 0.056)\,M_{\odot}$,
 where $N_{g}$ and $\langle M_{star} \rangle$ are, respectively, the number and the average mass of galaxies
 with $M_{star} > 10^{11}\,M_{\odot}$, and
 $f$ is the ratio of the mass of galaxies with $M>10^{11}\,M_{\odot}$ to 
 those with $M<10^{11}\,M_{\odot}$. In our case, we have $N_{g} = 6$ and
 $\langle M_{star} \rangle =5 \times 10^{11}\,M_{\odot}$.
  We adopt $f \sim 4$, which is the ratio of the luminosity density of
  $L > L_{*}$ galaxies to that of $L < L_{*}$ galaxies in the Schechter function
  with faint-end slope $\alpha \simeq$-1.24 (e.g, Rines \& Geller 2008).
  This gives an overdensity mass of (2.1-4.6)$\times 10^{14}\,M_{\odot}$

  It is interesting that several independent mass estimates give the overdensity
 mass to be $M \sim$ a few $\times\,10^{14}\,M_{\odot}$. The implausible
 assumption of the virialization even gives a mass consistent with
 other estimates. This suggests that the virialization of the proto-cluster may 
 soon be completed. However, the expected number density of halos as massive
 as a few $\times\,10^{14}\,M_{\odot}$ at $z\simeq 3.5$ is extremely low with
 $10^{-10}$ Mpc$^{-3}$ \citep{pk07,ree07}, and the probability of finding such a 
 halo in the CDF-S is only $10^{-5}$.
  This structure can be a part of the filaments and walls at high redshift as well.
  The existence of another weak overdense area such as the B overdensity supports
 such an idea. A careful analysis of simulation 
 data should help us understand the nature of these kinds of overdensities.

\section{Conclusions}
   By examining the two-dimensional distribution of
 $3.45 \lesssim z \lesssim 3.85$ candidates
 in the CDF-S field selected from the photometric redshift method or 
 the $BVz$ color-color space, we find plausible associations of the overdense
 region of such objects with AGNs and LBGs at $z_{spec} \sim 3.7$.
  The significance of the overdense regions
 is found to be $5-7\sigma$. The overdense area is abundant with massive
 galaxies, adding further support that this is a proto-cluster.
  The derived mass of the proto-cluster is found to be
 a few $\times$ $10^{14}$ $M_{\odot}$. 
  The existence of the massive proto-cluster in the CDF-S shows
 that one must be  
 cautious when examining the significance of high redshift overdensities in 
 other fields with respect to the CDF-S.
%\\
%\\
%\\
\acknowledgments NIR data come from the ESO La Silla and Paranal Observatories 
 under Program ID(s): LP168.A-0485, and 171.A-3045.
 This work was supported by the Creative Research Initiatives
 program, CEOU of MEST/KOSEF.
 We thank Ranga-Ram Chary for providing the IRAC photometry of objects in
 the CDF-S and for useful discussions.

\clearpage

\end{document}